\documentclass[9pt]{extarticle}

\usepackage{amsmath}
\usepackage{amssymb}
\usepackage[textstyle,amssymb]{SIunits}
\usepackage{graphicx}
\usepackage{fixltx2e}  
\usepackage[normalem]{ulem}  
\usepackage{tikz,pgfplots}   

\usepackage{color}
\usepackage{soul}

\usepackage[twocolumn,textwidth=183mm,columnsep=6mm,top=20mm,bottom=25mm]{geometry}
\usepackage{helvet}
\usepackage{titlesec}
\titleformat*{\section}{\bfseries\sffamily}
\titlespacing{\section}{0pt}{*4}{*0}
\titleformat{\subsection}[runin]{\normalfont\bfseries}{\thesubsection.}{3pt}{}
\usepackage{setspace}
\usepackage[breaklinks=false]{hyperref}

\usepackage[font={footnotesize,sf},labelfont={sf,bf},labelsep=endash]{caption}

\usepackage{natbib}
\usepackage{textcomp} 

\usepackage[hang,flushmargin]{footmisc}
\usepackage{pgf}

\begin{document}

\twocolumn[\begin{@twocolumnfalse}

{\Huge\sf \textbf{A monolithic frequency comb platform based on interband cascade lasers and detectors}}
		
\vspace{0.5cm}

{\sf\large \textbf {Benedikt~Schwarz$^{1*}$, Johannes~Hillbrand$^1$, Maximilian~Beiser$^1$, Aaron~Maxwell~Andrews$^{2}$, Gottfried~Strasser$^{2}$, Hermann~Detz$^{2,3}$, Anne Schade$^4$, Robert Weih$^{4,5}$ and Sven~H\"ofling$^{4,6}$}}
\vspace{0.5cm}

{\sf \textbf{$^1$Institute of Solid State Electronics, TU Wien, Gu{\ss}hausstra{\ss}e 25, 1040 Vienna, Austria\\
		$^2$Center for Micro- and Nanostructures, TU Wien, Gu{\ss}hausstra{\ss}e 25, 1040 Vienna, Austria\\
		$^3$Central European Institute of Technology, Brno University of Technology, Brno, Czech Republic\\
		$^4$Technische Physik, Physikalisches Institut, University W\"urzburg, Am Hubland, 97074 W\"urzburg, Germany\\
		$^5$Nanoplus Nanosystems and Technologies GmbH, 97218 Gerbrunn, Germany\\
		$^6$SUPA, School of Physics and Astronomy, University of St Andrews, St Andrews, KY16 9SS, United Kingdom\\
$^*$e-mail: {benedikt.schwarz@tuwien.ac.at}}}
\vspace{0.5cm}

\end{@twocolumnfalse}]


{\sf \small \textbf{\boldmath
		Optical frequency combs enable all-solid-state spectrometers that will trigger a breakthrough in miniaturization and on-chip integration of mid-infrared sensing technology.
		Interband cascade lasers (ICLs) are an ideal candidate for practical implementations due to their low power consumption and zero-bias detection functionality.
		Here, we demonstrate the generation of self-starting ICL frequency combs. 
		We show that the gain is fast enough to respond to beating of the intracavity field, which hinders the formation of short pulses. Instead, the ICL operates in a comb state, where the phases of the fundamental intermode beatings are splayed over a range of $2\pi$.
		This state appears to be general to self-starting combs based on the inherent gain nonlinearity and minimizes oscillations of the population inversion. Using the same epi-layer material, we demonstrate efficient detector operation at room temperature up to several GHz bandwidth and thereby provide a complete and unmatched platform for monolithic and battery driven dual-comb spectroscopy.
}}

Sensors are the heart of every smart technology. They allow to capture data of environmental pollution, plant infections, or our current physiological condition.
The mid-infrared is the spectral region of choice when it comes to sensing and spectroscopy. No other spectral region provides the same sensitivity or selectivity for molecular fingerprinting.
Established mid-infrared spectrometers are mostly based on free space optics and moving parts.
None of these concepts can be sufficiently down-scaled to single-chip dimensions. 
Dual-comb spectroscopy\cite{schiller2002spectrometry,keilmann2004time,coddington2016dual,ycas2018high} offers the unique possibility to directly map the optical spectrum to the electrical domain.
Dual-comb spectrometers do not require movable parts and can be miniaturized without losing spectral resolution -- there is no relation between the spectral resolution and the spatial dimension.
Its full potential for practical applications can be unlocked, if both frequency comb generators and heterodyne detectors are miniaturized, driven by a battery and ideally integrated on a chip.

For a long time, frequency comb generation in the mid-infrared was limited to table-top instruments\cite{reid2008frequency,andriukaitis2011GW}. 
A first compact implementation was demonstrated utilizing the Kerr nonlinearity in passive microresonators\cite{delHaye2007optical,kippenberg2011microresonator}. 
While microresonator combs recently reached the maturity to enable compact and battery driven systems at telecom wavelengths\cite{stern2018battery}, frequency comb generation in the mid-infrared is still restricted to external table-top pump lasers.

An alternative concept employs the third-order nonlinearity of the laser itself.
This per se monolithic approach is ideal for the realization of miniaturized spectrometers and was found in quantum cascade lasers (QCLs) a few years ago\cite{hugi2012mid}.
Due to the ultra-fast gain dynamics, the population inversion can respond to beatings between laser modes. The resulting four-wave mixing is responsible for coupling the modes and the formation of the frequency comb.
While QCL frequency combs are compact, their high power consumption strongly limits their use in battery driven applications.
Back in 1988, Agrawal pointed out that the four-wave mixing process does not necessarily require ultra-fast population inversion dynamics\cite{agrawal1988population}.
This suggests that the concept of self-starting frequency combs based on the inherent gain nonlinearity can also be transfered to other types of lasers that fulfill the requirements for portable devices.
ICLs are the ideal alternative\cite{yang1996novel,vurgaftman2015interband,vurgaftman2011rebalancing}, because their power consumption is 1--2 orders of magnitude smaller than that of QCLs.
Also in terms of output power and efficiency, ICLs are catching up rapidly\cite{vurgaftman2015interband,kim2015high,bradshaw1999high,canedy2015interband}.

\begin{figure*}[t]
	\centering
	\includegraphics[width=\textwidth]{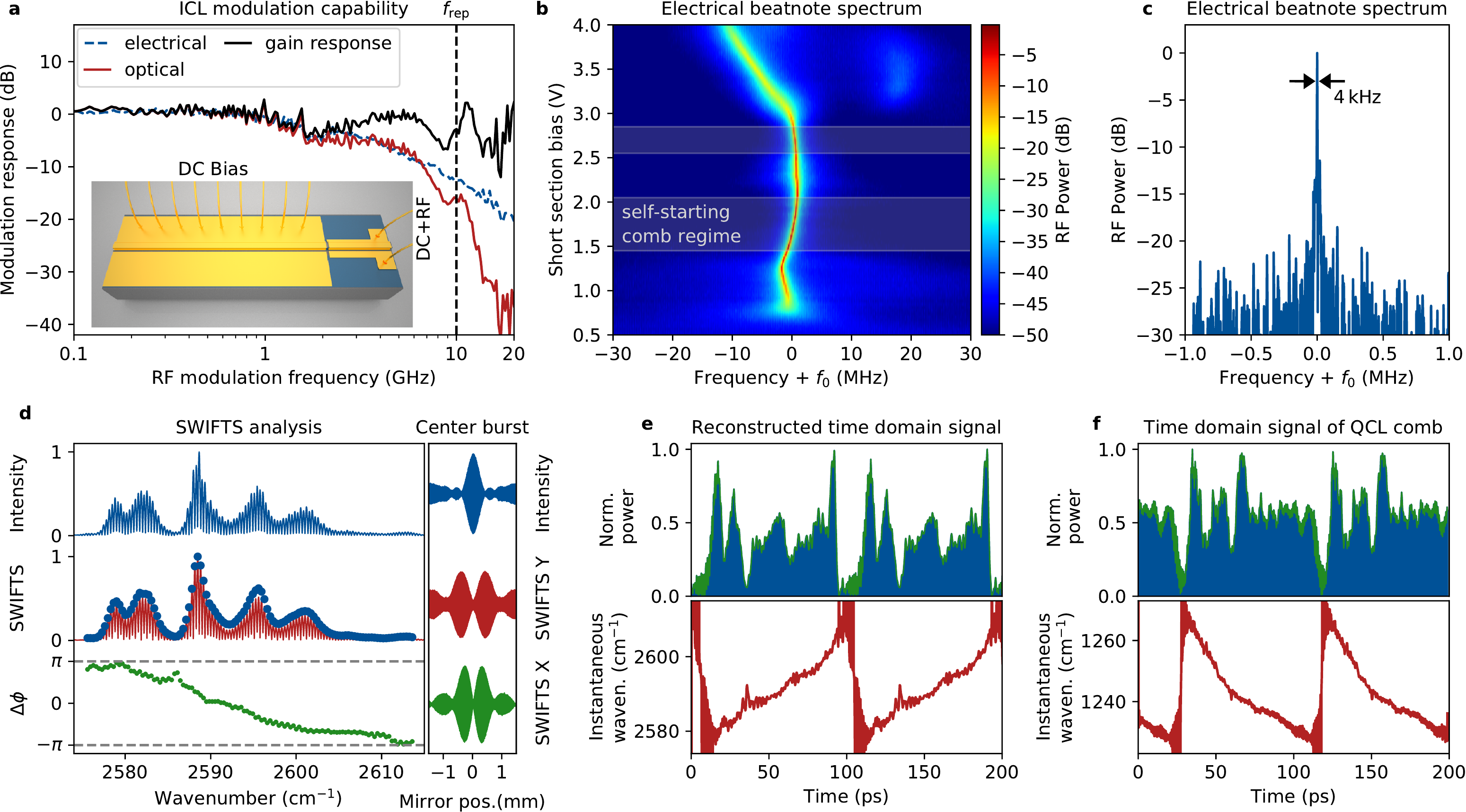}	
	\caption{\textbf{ICL frequency comb.} \textbf{a} Frequency modulation capabilities of the optimized devices measured optically via a fast QWIP (red) and electrically via RF rectification (blue). The black curve is corrected by the RC-cutoff of the ICL and QWIP response to extract the portion due to the response of the gain medium. Inset: sketch of the two-section device.
	\textbf{b} The spectral lines of the ICL beat together leading to a modulation of the laser intensity at the cavity roundtrip frequency. The resulting electrical beatnote can be measured by recording the RF spectrum of the absorber current and tunes with the absorber bias. Regions of self-starting comb operation are highlighted by the rectangles.
	\textbf{c} Narrow electrical beatnote with a linewidth of 4$\,\kilo\hertz$.
	\textbf{d{\&}e} '\textit{Shifted Wave Intermode Beat Fourier Transform Spectroscopy}' (SWIFTS) analysis and corresponding time domain signal of the ICL frequency comb. The intermode difference phases cover the range from $\pi$ to $-\pi$. This and the minimum in the SWIFTS interferograms at zero-phase are characteristic for the suppression of amplitude modulation.  \textbf{f} Time domain signal of a QCL frequency comb\cite{hillbrand2018coherent}. Irrespective of the sign, the both combs show the same characteristic dominantly frequency modulated outout. This suggests that the ICL frequency comb is indeed governed by the inherent gain nonlinearity, which is connected to the suppression of amplitude modulations.}
	\label{fig1}
\end{figure*}

In this work, we present a new monolithic frequency comb sensing platform based on ICLs.
We demonstrate self-starting frequency comb operation of ICLs based on the inherent gain nonlinearity.
Despite the fact that the laser transition lifetime differs by more than two orders of magnitude, ICLs can respond to beatings between laser modes and the observed phase pattern is strikingly similar to QCLs. Our experiments reveal that ICL frequency combs are characterized by a strong suppression of amplitude modulation.
We further show that ICL combs can be locked to an external RF oscillator while maintaining full intermodal coherence. This allows an all-electric control and stabilization of the frequency comb against the harsh conditions, which are omnipresent in real-life applications.
Finally, we highlight the unique detection functionality naturally provided by the ICL material, which can be used to directly integrate sensitive multi-heterodyne detectors.
Hence, the ICL based monolithic platform provides all features to realize miniaturized and battery driven dual-comb spectrometers.

\begin{figure*}
	\centering
	\includegraphics[width=0.99\linewidth]{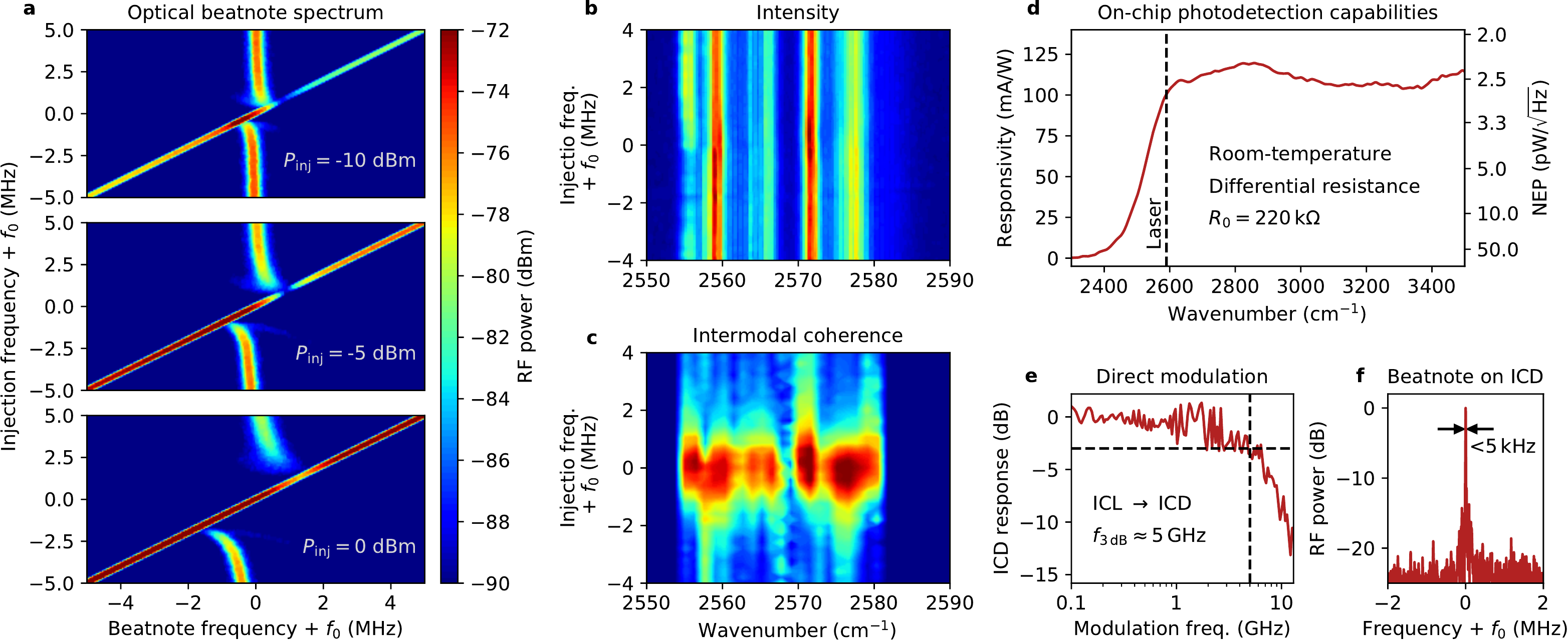}
	\caption{\textbf{Coherent injection locking and on-chip detection capabilities.}
		\textbf{a} Optical beatnote spectrum of the ICL depending on the injection frequency for three different injection powers. The large locking range is due to the optimized RF injection.
		\textbf{b{\&}c} Intensity and SWIFTS spectral maps at 5dB injection power as function of the injection frequency.
		\textbf{d} Spectral responsivity and noise equivalent power of the on-chip interband cascade detector (ICD), which was cleaved from the same chip. \textbf{e} Direct modulation response of the ICL measured with the ICD. \textbf{f} Narrow beatnote of the ICL comb measured with the ICD.}
	\label{fig2}
\end{figure*}

In order to investigate frequency comb operation of ICLs, we fabricated a two section device (fig.~\ref{fig1}a).
The short section is optimized for efficient radio-frequency (RF) extraction and injection and its DC bias can be adjusted to alter the locking properties\cite{yang2017achieving}.
The high frequency modulation response provides important information about the internal gain dynamics.
The extracted frequency response is flat up to 20$\,$GHz and reveals that the characteristic response time of the population inversion $\tau_\mathrm{\Delta n}$ exceeds at least the first two harmonics of the repetition frequency.
Since the condition $\tau_\mathrm{\Delta n}f_\mathrm{rep} \lesssim 1$ is satisfied, four-wave mixing via population inversion oscillations is the dominant contribution to the gain nonlinearity\cite{agrawal1988population} and enables self-starting frequency comb operation.

A first indication for frequency comb operation is the observation of a narrow beatnote. 
Each pair of laser lines causes a beating at their difference frequency. In frequency combs all lines are equidistant, resulting in a narrow beatnote.
Fig.~\ref{fig1}b shows the electrically extracted laser beatnote depending on the short section bias. The latter can be altered to tune the ICL into the self-locking regime.
There, the beatnote appears narrow with a linewidth on the kHz level (fig.~\ref{fig1}c), which is more than an order of magnitude below the Shawlow-Townes quantum limited linewidth of the unlocked laser.

In order to unequivocally proof frequency comb operation we employ SWIFTS\cite{burghoff2014terahertz,burghoff2015evaluating}. This technique provides a complete characterization of a comb state by measuring the coherence and phase between each pair of comb lines.
The SWIFTS analysis of the ICL frequency comb is plotted in fig.~\ref{fig1}d. The amplitudes of the SWIFTS spectrum match the beating amplitudes extracted from the intensity spectrum (blue dots), which serves as a proof for comb operation over the entire emission spectrum. The minimum of the SWIFTS interferograms at zero path difference is a clear sign of the suppression of amplitude modulations. Even more exciting is additional information provided by the phases. The intermodal phases follow a chirped pattern covering a range of $2\pi$. The same pattern was recently observed in QCLs\cite{singleton2018evidence}.
Fig.~\ref{fig1}e{\&}f show the direct comparison of the time domain signals of an ICL and a QCL frequency comb.
Although the laser transition lifetimes differ by two orders of magnitude and operate at a different wavelength, both frequency combs show the linearly chirped instantaneous frequency accompanied by the suppression of amplitude modulations.
This remarkable similarity indicates that the comb formation is indeed governed by the same physical mechanism.
A state with a weak amplitude modulation maximized the round-trip gain, as amplitude modulations lead to a stronger gain saturation in fast gain media. This is the opposite of passive mode-locking in slow gain media, where four-wave mixing occurs in the fast saturable absorption, whereas the gain section is too slow to respond to intermode beatings.
There, the formation of pulses is advantageous, as it minimizes the round-trip loss.

Recent findings revealed that the repetition frequency of a QCL frequency comb can be synchronized to an external RF oscillator, while maintaining full intermodal coherence\cite{hillbrand2018coherent}. This increases the range of comb operation, enables all-electric frequency comb stabilization via control loops and increases the stability against optical feedback.
This technique can also be employed with ICL frequency combs.
Fig.~\ref{fig2}a shows the optically measured beatnote spectrum while sweeping the injection frequency across the beatnote for three different injection power levels. The laser beatnote is fully controlled by the external RF oscillator within a locking range of 1, 2 and 4$\,$MHz, respectively.
Fig.~\ref{fig2}b shows that full intermodal coherence is achieved within the locking range, which serves as a proof that ICL frequency combs can be electrically injection locked.

Integrating a complete dual-comb spectrometer on a chip requires a technology that is capable of both frequency comb operation and high speed multi-heterodyn detection.
An elegant solution is building the laser and detector from the very same material. 
This so-called bi-functional operation of the very same material was first demonstrated using QCLs\cite{schwarz2012bi} and trigged the realization of a monolithic lab-on-a-chip\cite{schwarz2014monolithically}. However, the additional on-chip detection feature requires large modifications of the QCL design to match the laser and detector wavelength. Hence, bi-functional QCLs were restricted to pulsed operation for a long time\cite{schwarz2017watt}.
Fortunately the bi-functional operation of ICLs comes naturally and does not require modifications of the design\cite{lotfi2016monolithically}.
The longer lifetime of the optical transition decreases the thermal noise and increases the internal quantum efficiency.
Fig.~\ref{fig2}d shows the spectral responsivity of the interband cascade detector (ICD) that was fabricated on the same chip and cleaved for the characterization.
 A broadband spectral responsivity of over $100\,$mA/W and an exceptional noise equivalent power of $2.5\,\pico\watt\per\sqrt{\hertz}$ are achieved at room temperature. 
The provided, yet not optimized, ICD performance is clearly superior to bi-functional QCL material.
In order to demonstrate the high-speed operation, which is essential for multi-heterodyne detection, we measured the ICD frequency response optically using the ICL (fig.~\ref{fig2}e). The high frequency cut-off around $5\,$GHz includes both the laser and the detector response and is limited by parasitic capacitances. 
The large detection bandwidth even enables the measurement of the beatnote of the ICL frequency comb with an SNR of 22dB at room temperature (fig.~\ref{fig2}f).
Surprisingly, this is even higher than the SNR obtained with a cryogenically cooled 7$\,$GHz RF-QWIP. These results show the remarkable detection capabilities provided by the ICL comb platform. Integrated on the same chip, the coupling efficiencies exceed free-space optics by at least two orders of magnitude.
In principle, a monolithic sensor chip with a few centimeters of interaction length can thus achieve the same detector noise limited sensitivity as a setup with discrete elements and an interaction length of a few meters.

\begin{figure}[h!]
	\begin{normalfont}
		\begin{tabular}{ l c c c }
			\textbf{Requirements} & \textbf{Microresonator} & \textbf{QCL} & \textbf{ICL}\\\hline 
			\noalign{\vskip 1mm} 
			comb operation  & $\circ$ & $\circ$ & $\bullet$ \\
			stabilization knobs  & $\circ$ & $\circ$ & $\bullet$ \\
			compact & $\boldsymbol{-}^*$  & $\circ$ & $\bullet$ \\
			battery driven & $\boldsymbol{-}^*$ & $\boldsymbol{-}$ & $\bullet$ \\
			GHz on-chip detection & $\boldsymbol{-}$  & $\boldsymbol{-}$ & $\bullet$\\
		\end{tabular}
		
		\vspace{0.1cm}
		(-) not demonstrated, ($\circ$) demonstrated, ($\bullet$) this work.
		
		$^*$ Only in the near-infrared\cite{stern2018battery}.
	\end{normalfont}
	\captionof{table}{The presented ICL frequency comb platform is the only available mid-infrared technology that meets all requirements to build ultra-compact and battery driven dual-comb spectrometers.}
	\label{table1}
\end{figure}

In conclusion, we presented a new mid-infrared frequency comb platform based on ICLs. In contrast to other technologies, the ICL comb platform provides all properties for ultra-compact and battery driven dual-comb sensors, as summarized in table~\ref{table1}.
We demonstrated self-starting ICL frequency combs based on the inherent gain nonlinearity caused by the fast dynamics of the population inversion.
The resulting frequency comb state shows a dominant frequency modulation and a linearly chirped phase signature, which is characteristic for the suppression of amplitude modulations in fast gain media.
ICL technology reduces the power consumption by two orders of magnitude compared to QCL to below one watt, enabling battery driven sensors.
Furthermore, ICLs provide exceptionally sensitive and high speed photodetection capabilities that allow monolithic integration of all active optical components. 
Thus, ICL technology provides all  properties for all-solid-state mid-infrared sensors and will promote dual-comb spectroscopy from fundamental research to a broadly used sensing platform.

\newpage
\section*{Methods}
\footnotesize
\setstretch{1.}

\subsection*{Device fabrication}
The investigated ICLs were grown at the Universit\"at W\"urzburg and processed at the Center of Micro- and Nanostructures at TU Wien. The simulated bandstructure is shown in supp. fig.~6.
The laser ridges were fabricated with standard mask photolithography, reactive ion etching to define the laser waveguides, silicon nitride for the passivation layers and sputtered TiAu for the contact pads. The substrate was thinned to 160$\,\micro\meter$ and TiAu was sputtered on the backside  . The devices were mounted epi-side up using indium on a copper mount. Different thickness of the passivation layer have been used to optimize the device performance in terms of output power and modulation capability.

\vspace{-0.3cm}
\subsection*{Measurement setup}
The devices were mounted on a copper submount, thermo-electrically stabilized to 15 $^{\circ}$C. In order to improve the noise properties of our laser driver, we used home-built low-pass filters. The accuracy of temperature stabilization and the noise of the laser drivers play a crucial role for the stability of the frequency comb.

\vspace{-0.3cm}
\subsection*{Relevance of the spatial beatnote profile}
An important aspect has to be considered for injection locking, is the spatial patterns of the laser beatnote in standing wave lasers\cite{piccardo2018time}. Because of the boundary condition of the modes in the cavity, all beatings between adjacent lines will follow a half wave cosine function along the cavity. Hence, the modulation should be applied at the end of the cavity, where the laser is most susceptible to injection locking.

\vspace{-0.3cm}
\subsection*{SWIFTS}
The SWIFTS concept was realized using a fast QWIP placed at the output window of a FTIR (sketch in supp. fig.~5). The emitted light of the ICL is shined through the FTIR onto the QWIP. A local oscillator mixes down the the optical beatnote of the ICL to $\approx$ 40 MHz. By recording the quadrature components X and Y of the QWIP signal in dependence of the delay time $\tau$ using a lock-in amplifier, we obtain the two SWIFTS quadrature interferograms. The reference signal for the lock-in amplifier is obtained by mixing another local oscillator with the RF source used for injection into the ICL. All interferograms were recorded by the lock-in using the Helium-Neon trigger of the FTIR. Our FTIR (Bruker Vertex 70v) moves each arm by $\pm \tau /2$. The complex sum of the interferograms is therefore given by
\begin{equation}
\mathbf{F}(X+iY)(\tau)= \sum_n A_n A_{n-1}  \left[ \cos \left( \frac{\omega_r \tau}{2} \right) + \cos \left(\omega_{n-1/2} \tau \right) \right].
\end{equation}
By applying the fast fourier transformation, using zero-padding and peak fitting, both amplitude and phase of all intermode beatings are obtained. The intermode beatings describe the coherence and phases between adjacent comb modes.
The normalized intermodal coherence can be defined as~\cite{burghoff2015evaluating}. 
\begin{equation}
c=\frac{|<A_nA_{n-1}e^{(i(\phi_n-\phi_{n-1}))}>|}{<|A_n||A_{n-1}|>}, \label{eq:coherence}
\end{equation} 
where $A_n$ are the field amplitudes, $\phi_n$ the intermodal phases and the brackets denote temporal averaging.
The phases of the modes can be calculated by the cumulative sum of the intermode beating phases, which together with the intensity spectrum allows the reconstruction of the time domain signal of a comb state.



\footnotesize
\setstretch{1.}

\providecommand{\noopsort}[1]{}\providecommand{\singleletter}[1]{#1}%

\renewcommand\refname{\normalsize References}

\providecommand{\noopsort}[1]{}\providecommand{\singleletter}[1]{#1}%

\footnotesize

\section*{Acknowledgements}
This work was supported by the Austrian Science Fund (FWF) within the projects "NanoPlas" (P28914-N27), "Building Solids for Function" (Project W1243), "NextLite" (F4909-N23), as well as by the Austrian Research Promotion Agency through the ERA-Net Photonic Sensing program, project "ATMO-SENSE" (FFG: 861581). H.D. was supported by the ESF project CZ.02.2.69/0.0/ 0.0/16\_027/0008371. A.M.A was supported by the projects COMTERA - FFG 849614 and AFOSR FA9550-17-1-0340. B.~Schwarz thanks A.~Belyanin for the fruitful discussions at IQCLSW 2018.

\section*{Author contributions}
B.S. planned and organized this project and wrote the manuscript. J.H. and B.S. performed the phase measurements of the combs. J.H. fabricated and tested the RF QWIPs. M.B. designed, fabricated and tested the two-section ICLs with guidance of B.S.. H.D., A.M.A and G.S. grew the QWIP structures. A.S., R.W. and S.H. provided the ICL material. All authors contributed to discussions how to interpret the results and commented on the manuscript.

\section*{Competing financial interests}
The authors declare no competing financial interests.

\section*{Data availability statement}
The data that support the plots within this paper and other findings of this study are available from the corresponding author upon reasonable request.


\begin{thebibliography}{10}
	\expandafter\ifx\csname url\endcsname\relax
	\def\url#1{\texttt{#1}}\fi
	\expandafter\ifx\csname urlprefix\endcsname\relax\def\urlprefix{URL }\fi
	\providecommand{\bibinfo}[2]{#2}
	\providecommand{\eprint}[2][]{\url{#2}}
	
	\bibitem{schiller2002spectrometry}
	\bibinfo{author}{Schiller, S.}
	\newblock \bibinfo{title}{Spectrometry with frequency combs}.
	\newblock \emph{\bibinfo{journal}{Optics Letters}}
	\textbf{\bibinfo{volume}{27}}, \bibinfo{pages}{766} (\bibinfo{year}{2002}).
	\newblock \urlprefix\url{https://doi.org/10.1364/ol.27.000766}.
	
	\bibitem{keilmann2004time}
	\bibinfo{author}{Keilmann, F.}, \bibinfo{author}{Gohle, C.} \&
	\bibinfo{author}{Holzwarth, R.}
	\newblock \bibinfo{title}{Time-domain mid-infrared frequency-comb
		spectrometer}.
	\newblock \emph{\bibinfo{journal}{Optics Letters}}
	\textbf{\bibinfo{volume}{29}}, \bibinfo{pages}{1542} (\bibinfo{year}{2004}).
	\newblock \urlprefix\url{https://doi.org/10.1364/ol.29.001542}.
	
	\bibitem{coddington2016dual}
	\bibinfo{author}{Coddington, I.}, \bibinfo{author}{Newbury, N.} \&
	\bibinfo{author}{Swann, W.}
	\newblock \bibinfo{title}{Dual-comb spectroscopy}.
	\newblock \emph{\bibinfo{journal}{Optica}} \textbf{\bibinfo{volume}{3}},
	\bibinfo{pages}{414} (\bibinfo{year}{2016}).
	\newblock \urlprefix\url{https://doi.org/10.1364/optica.3.000414}.
	
	\bibitem{ycas2018high}
	\bibinfo{author}{Ycas, G.} \emph{et~al.}
	\newblock \bibinfo{title}{High-coherence mid-infrared dual-comb spectroscopy
		spanning 2.6 to 5.2{\hspace{0.167em}}$\upmu$m}.
	\newblock \emph{\bibinfo{journal}{Nature Photonics}}
	\textbf{\bibinfo{volume}{12}}, \bibinfo{pages}{202--208}
	(\bibinfo{year}{2018}).
	\newblock \urlprefix\url{https://doi.org/10.1038/s41566-018-0114-7}.
	
	\bibitem{reid2008frequency}
	\bibinfo{author}{Reid, D.~T.}, \bibinfo{author}{Gale, B. J.~S.} \&
	\bibinfo{author}{Sun, J.}
	\newblock \bibinfo{title}{Frequency comb generation and carrier-envelope phase
		control in femtosecond optical parametric oscillators}.
	\newblock \emph{\bibinfo{journal}{Laser Physics}}
	\textbf{\bibinfo{volume}{18}}, \bibinfo{pages}{87--103}
	(\bibinfo{year}{2008}).
	\newblock \urlprefix\url{https://doi.org/10.1134/s1054660x08020011}.
	
	\bibitem{andriukaitis2011GW}
	\bibinfo{author}{Andriukaitis, G.} \emph{et~al.}
	\newblock \bibinfo{title}{90 {GW} peak power few-cycle mid-infrared pulses from
		an optical parametric amplifier}.
	\newblock \emph{\bibinfo{journal}{Optics Letters}}
	\textbf{\bibinfo{volume}{36}}, \bibinfo{pages}{2755} (\bibinfo{year}{2011}).
	\newblock \urlprefix\url{https://doi.org/10.1364/ol.36.002755}.
	
	\bibitem{delHaye2007optical}
	\bibinfo{author}{Del'Haye, P.} \emph{et~al.}
	\newblock \bibinfo{title}{Optical frequency comb generation from a monolithic
		microresonator}.
	\newblock \emph{\bibinfo{journal}{Nature}} \textbf{\bibinfo{volume}{450}},
	\bibinfo{pages}{1214--1217} (\bibinfo{year}{2007}).
	\newblock \urlprefix\url{https://doi.org/10.1038/nature06401}.
	
	\bibitem{kippenberg2011microresonator}
	\bibinfo{author}{Kippenberg, T.~J.}, \bibinfo{author}{Holzwarth, R.} \&
	\bibinfo{author}{Diddams, S.~A.}
	\newblock \bibinfo{title}{Microresonator-based optical frequency combs}.
	\newblock \emph{\bibinfo{journal}{Science}} \textbf{\bibinfo{volume}{332}},
	\bibinfo{pages}{555--559} (\bibinfo{year}{2011}).
	\newblock \urlprefix\url{https://doi.org/10.1126/science.1193968}.
	
	\bibitem{stern2018battery}
	\bibinfo{author}{Stern, B.}, \bibinfo{author}{Ji, X.},
	\bibinfo{author}{Okawachi, Y.}, \bibinfo{author}{Gaeta, A.~L.} \&
	\bibinfo{author}{Lipson, M.}
	\newblock \bibinfo{title}{Battery-operated integrated frequency comb
		generator}.
	\newblock \emph{\bibinfo{journal}{Nature}} \textbf{\bibinfo{volume}{562}},
	\bibinfo{pages}{401--405} (\bibinfo{year}{2018}).
	\newblock \urlprefix\url{https://doi.org/10.1038/s41586-018-0598-9}.
	
	\bibitem{hugi2012mid}
	\bibinfo{author}{Hugi, A.}, \bibinfo{author}{Villares, G.},
	\bibinfo{author}{Blaser, S.}, \bibinfo{author}{Liu, H.~C.} \&
	\bibinfo{author}{Faist, J.}
	\newblock
	\bibinfo{title}{\href{https://doi.org/10.1038/nature11620}{Mid-infrared
			frequency comb based on a quantum cascade laser}}.
	\newblock \emph{\bibinfo{journal}{Nature}} \textbf{\bibinfo{volume}{492}},
	\bibinfo{pages}{229--233} (\bibinfo{year}{2012}).
	
	\bibitem{agrawal1988population}
	\bibinfo{author}{Agrawal, G.~P.}
	\newblock
	\bibinfo{title}{\href{https://doi.org/10.1364/josab.5.000147}{Population
			pulsations and nondegenerate four-wave mixing in semiconductor lasers and
			amplifiers}}.
	\newblock \emph{\bibinfo{journal}{Journal of the Optical Society of America B}}
	\textbf{\bibinfo{volume}{5}}, \bibinfo{pages}{147} (\bibinfo{year}{1988}).
	
	\bibitem{yang1996novel}
	\bibinfo{author}{Yang, R.~Q.} \& \bibinfo{author}{Pei, S.~S.}
	\newblock \bibinfo{title}{Novel type-{II} quantum cascade lasers}.
	\newblock \emph{\bibinfo{journal}{Journal of Applied Physics}}
	\textbf{\bibinfo{volume}{79}}, \bibinfo{pages}{8197--8203}
	(\bibinfo{year}{1996}).
	\newblock \urlprefix\url{https://doi.org/10.1063/1.362554}.
	
	\bibitem{vurgaftman2015interband}
	\bibinfo{author}{Vurgaftman, I.} \emph{et~al.}
	\newblock \bibinfo{title}{Interband cascade lasers}.
	\newblock \emph{\bibinfo{journal}{Journal of Physics D: Applied Physics}}
	\textbf{\bibinfo{volume}{48}}, \bibinfo{pages}{123001}
	(\bibinfo{year}{2015}).
	\newblock \urlprefix\url{http://stacks.iop.org/0022-3727/48/i=12/a=123001}.
	
	\bibitem{vurgaftman2011rebalancing}
	\bibinfo{author}{Vurgaftman, I.} \emph{et~al.}
	\newblock \bibinfo{title}{Rebalancing of internally generated carriers for
		mid-infrared interband cascade lasers with very low power consumption}.
	\newblock \emph{\bibinfo{journal}{Nature Communications}}
	\textbf{\bibinfo{volume}{2}} (\bibinfo{year}{2011}).
	\newblock \urlprefix\url{https://doi.org/10.1038/ncomms1595}.
	
	\bibitem{kim2015high}
	\bibinfo{author}{Kim, M.} \emph{et~al.}
	\newblock \bibinfo{title}{High-power continuous-wave interband cascade lasers
		with 10 active stages}.
	\newblock \emph{\bibinfo{journal}{Optics Express}}
	\textbf{\bibinfo{volume}{23}}, \bibinfo{pages}{9664} (\bibinfo{year}{2015}).
	\newblock \urlprefix\url{https://doi.org/10.1364/oe.23.009664}.
	
	\bibitem{bradshaw1999high}
	\bibinfo{author}{Bradshaw, J.~L.}, \bibinfo{author}{Yang, R.~Q.},
	\bibinfo{author}{Bruno, J.~D.}, \bibinfo{author}{Pham, J.~T.} \&
	\bibinfo{author}{Wortman, D.~E.}
	\newblock \bibinfo{title}{High-efficiency interband cascade lasers with peak
		power exceeding 4 w/facet}.
	\newblock \emph{\bibinfo{journal}{Applied Physics Letters}}
	\textbf{\bibinfo{volume}{75}}, \bibinfo{pages}{2362--2364}
	(\bibinfo{year}{1999}).
	\newblock \urlprefix\url{https://doi.org/10.1063/1.125015}.
	
	\bibitem{canedy2015interband}
	\bibinfo{author}{Canedy, C.~L.} \emph{et~al.}
	\newblock \bibinfo{title}{Interband cascade lasers with $>$40{\%}
		continuous-wave wallplug efficiency at cryogenic temperatures}.
	\newblock \emph{\bibinfo{journal}{Applied Physics Letters}}
	\textbf{\bibinfo{volume}{107}}, \bibinfo{pages}{121102}
	(\bibinfo{year}{2015}).
	\newblock \urlprefix\url{https://doi.org/10.1063/1.4931498}.
	
	\bibitem{hillbrand2018coherent}
	\bibinfo{author}{Hillbrand, J.}, \bibinfo{author}{Andrews, A.~M.},
	\bibinfo{author}{Detz, H.}, \bibinfo{author}{Strasser, G.} \&
	\bibinfo{author}{Schwarz, B.}
	\newblock \bibinfo{title}{Coherent injection locking of quantum cascade laser
		frequency combs}.
	\newblock \emph{\bibinfo{journal}{arXiv preprint arXiv:1808.06636}}
	(\bibinfo{year}{2018}).
	\newblock \urlprefix\url{https://arxiv.org/abs/1808.06636}.
	
	\bibitem{yang2017achieving}
	\bibinfo{author}{Yang, Y.}, \bibinfo{author}{Burghoff, D.},
	\bibinfo{author}{Reno, J.} \& \bibinfo{author}{Hu, Q.}
	\newblock \bibinfo{title}{Achieving comb formation over the entire lasing range
		of quantum cascade lasers}.
	\newblock \emph{\bibinfo{journal}{Optics Letters}}
	\textbf{\bibinfo{volume}{42}}, \bibinfo{pages}{3888} (\bibinfo{year}{2017}).
	\newblock \urlprefix\url{https://doi.org/10.1364/ol.42.003888}.
	
	\bibitem{burghoff2014terahertz}
	\bibinfo{author}{Burghoff, D.} \emph{et~al.}
	\newblock
	\bibinfo{title}{\href{https://doi.org/10.1038/nphoton.2014.85}{Terahertz
			laser frequency combs}}.
	\newblock \emph{\bibinfo{journal}{Nature Photonics}}
	\textbf{\bibinfo{volume}{8}}, \bibinfo{pages}{462--467}
	(\bibinfo{year}{2014}).
	
	\bibitem{burghoff2015evaluating}
	\bibinfo{author}{Burghoff, D.} \emph{et~al.}
	\newblock
	\bibinfo{title}{\href{https://doi.org/10.1364/oe.23.001190}{Evaluating the
			coherence and time-domain profile of quantum cascade laser frequency combs}}.
	\newblock \emph{\bibinfo{journal}{Optics Express}}
	\textbf{\bibinfo{volume}{23}}, \bibinfo{pages}{1190} (\bibinfo{year}{2015}).
	
	\bibitem{singleton2018evidence}
	\bibinfo{author}{Singleton, M.}, \bibinfo{author}{Jouy, P.},
	\bibinfo{author}{Beck, M.} \& \bibinfo{author}{Faist, J.}
	\newblock \bibinfo{title}{Evidence of linear chirp in mid-infrared quantum
		cascade lasers}.
	\newblock \emph{\bibinfo{journal}{Optica}} \textbf{\bibinfo{volume}{5}},
	\bibinfo{pages}{948} (\bibinfo{year}{2018}).
	\newblock \urlprefix\url{https://doi.org/10.1364/optica.5.000948}.
	
	\bibitem{schwarz2012bi}
	\bibinfo{author}{Schwarz, B.} \emph{et~al.}
	\newblock \bibinfo{title}{A bi-functional quantum cascade device for
		same-frequency lasing and detection}.
	\newblock \emph{\bibinfo{journal}{Applied Physics Letters}}
	\textbf{\bibinfo{volume}{101}}, \bibinfo{pages}{191109}
	(\bibinfo{year}{2012}).
	\newblock \urlprefix\url{https://doi.org/10.1063/1.4767128}.
	
	\bibitem{schwarz2014monolithically}
	\bibinfo{author}{Schwarz, B.} \emph{et~al.}
	\newblock \bibinfo{title}{Monolithically integrated mid-infrared lab-on-a-chip
		using plasmonics and quantum cascade structures}.
	\newblock \emph{\bibinfo{journal}{Nature Communications}}
	\textbf{\bibinfo{volume}{5}} (\bibinfo{year}{2014}).
	\newblock \urlprefix\url{https://doi.org/10.1038/ncomms5085}.
	
	\bibitem{schwarz2017watt}
	\bibinfo{author}{Schwarz, B.} \emph{et~al.}
	\newblock \bibinfo{title}{Watt-level continuous-wave emission from a
		bifunctional quantum cascade laser/detector}.
	\newblock \emph{\bibinfo{journal}{{ACS} Photonics}}
	\textbf{\bibinfo{volume}{4}}, \bibinfo{pages}{1225--1231}
	(\bibinfo{year}{2017}).
	\newblock \urlprefix\url{https://doi.org/10.1021/acsphotonics.7b00133}.
	
	\bibitem{lotfi2016monolithically}
	\bibinfo{author}{Lotfi, H.} \emph{et~al.}
	\newblock \bibinfo{title}{Monolithically integrated mid-{IR} interband cascade
		laser and photodetector operating at room temperature}.
	\newblock \emph{\bibinfo{journal}{Applied Physics Letters}}
	\textbf{\bibinfo{volume}{109}}, \bibinfo{pages}{151111}
	(\bibinfo{year}{2016}).
	\newblock \urlprefix\url{https://doi.org/10.1063/1.4964837}.
	
	\bibitem{piccardo2018time}
	\bibinfo{author}{Piccardo, M.} \emph{et~al.}
	\newblock \bibinfo{title}{Time-dependent population inversion gratings in laser
		frequency combs}.
	\newblock \emph{\bibinfo{journal}{Optica}} \textbf{\bibinfo{volume}{5}},
	\bibinfo{pages}{475} (\bibinfo{year}{2018}).
	\newblock \urlprefix\url{https://doi.org/10.1364/optica.5.000475}.
	
\end{thebibliography}
\end{document}


\onecolumn

	{\noindent \Huge\sf \textbf{Supplementary information -- }}
	{\noindent \Huge\sf \textbf{A monolithic mid-infrared frequency comb platform based on interband cascade lasers}}
	
	\vspace{0.5cm}
	{\noindent \sf\large \textbf {Benedikt~Schwarz$^{1*}$, Johannes~Hillbrand$^1$, Maximilian~Beiser$^1$, Aaron~Maxwell~Andrews$^{2}$, Gottfried~Strasser$^{2}$, Hermann~Detz$^{2,3}$, Anne Schade$^4$, Robert Weih$^{4,5}$ and Sven~H\"ofling$^{4,6}$}}
	\vspace{0.5cm}
	
	\let\thefootnote\relax\footnotetext{
		 \sf {\noindent $^1$Institute of Solid State Electronics, TU Wien, Gu{\ss}hausstra{\ss}e 25, 1040 Vienna, Austria}\\
			$^2$Center for Micro- and Nanostructures, TU Wien, Gu{\ss}hausstra{\ss}e 25, 1040 Vienna, Austria\\
			$^3$Central European Institute of Technology, Brno University of Technology, Brno, Czech Republic\\
			$^4$Technische Physik, Physikalisches Institut, University W\"urzburg, Am Hubland, 97074 W\"urzburg, Germany\\
			$^5$Nanoplus Nanosystems and Technologies GmbH, 97218 Gerbrunn, Germany\\
			$^6$SUPA, School of Physics and Astronomy, University of St Andrews, St Andrews, KY16 9SS, United Kingdom\\
			$^*$e-mail: {benedikt.schwarz@tuwien.ac.at}}

\begin{figure*}[h]
	\centering
	\includegraphics[width=0.9\linewidth]{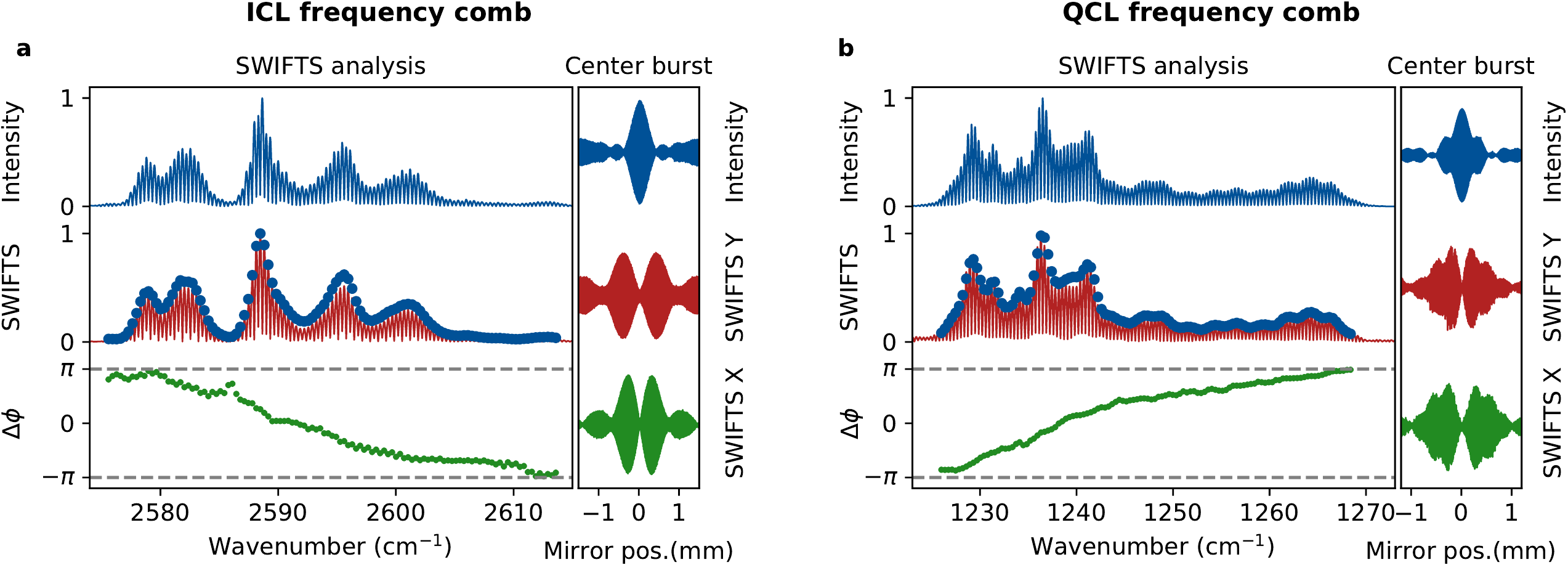}
	\caption{\textbf{Comparison of ICL and QCL frequency combs} \textbf{a} SWIFTS analysis of the ICL frequency comb, as shown in the paper. The intermode difference phases cover the range from $\pi$ to $-\pi$ from the lowest to the highest frequency mode of the optical spectrum. \textbf{b} SWIFTS analysis and corresponding time domain signal of a QCL frequency comb~[18]. The QCL frequency comb shows the same linear chirp of the intermode difference phases with the subtle difference of the positive slope. Irrespective of the sign of the slope, both states shown in this figure constitute a phase balancing state.}
\end{figure*}

\begin{figure}[h]
	\centering
	
	\begin{minipage}{0.5\textwidth}
		\includegraphics[width=0.9\linewidth]{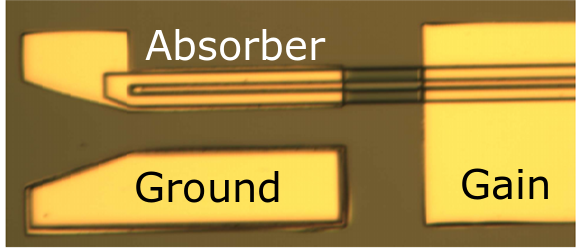}
	\end{minipage}%
	\begin{minipage}{0.35\textwidth}
		\includegraphics[width=0.9\linewidth]{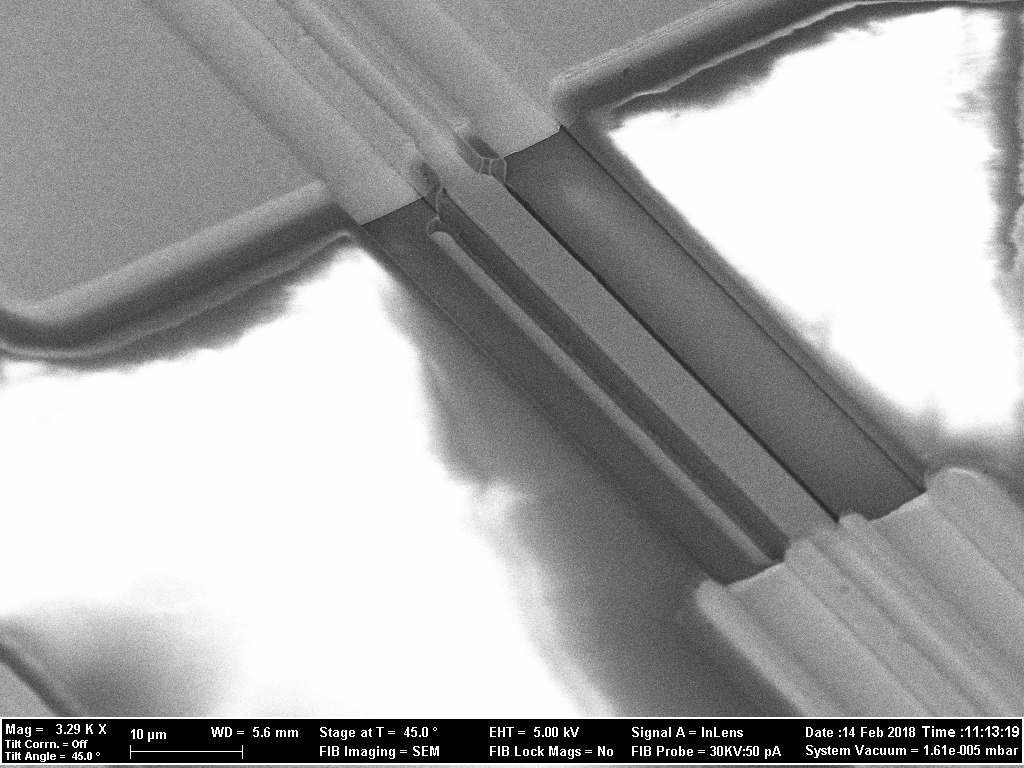}
	\end{minipage}

	\caption{\textbf{a}: Microscope image of a typical device. A 40 GHz signal-ground RF probe is positioned between the top contact of 160 \textmu m long absorber section and the ground pad beside it. The SiN passivation layer of the absorber section is chosen to be relatively thick in order to decrease the parasitic capacitance of the absorber top contact enabling efficient RF injection. \textbf{b}: Scanning electron microscope image of the section of the ridge waveguide between the absorber and the gain section. This section was 60 \textmu m long for all tested devices. This length was chosen to separate the top contacts of the absorber and gain sections electrically in order to allow different biasing of the two sections without permanently burning the device by excessive transverse current. Furthermore, the uncontacted section was also covered with a passivating SiN layer in order to ensure the continuity of the effective refractive index of the waveguide mode and prevent coupled cavity effects.}
\end{figure}

\begin{figure*}[h]
	\centering
	\includegraphics[width=0.9\linewidth]{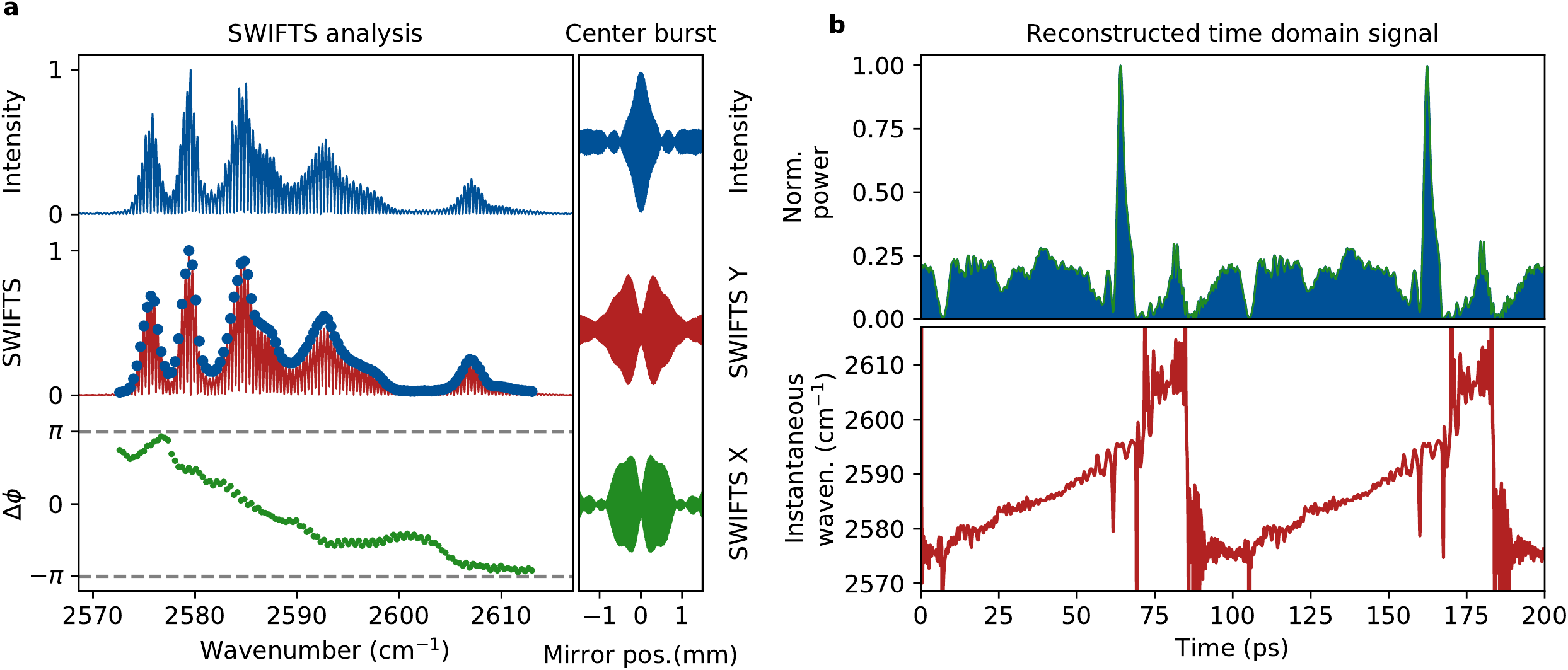}
	\caption{\textbf{Another repulsively synchronized frequency comb state:} \textbf{a}: The intensity spectrum spans over roughly 40 cm$^{-1}$. The SWIFTS spectrum has the same shape as the intensity spectrum without any spectral holes, proofing that the frequency comb is phase-coherent. The phases clearly show a phase balance state covering the range between $\pi$ and $-\pi$ due to repulsive synchronization. The center burst of the corresponding  SWIFTS interferograms (zoom) show a local minimum that is characteristic for FM-type comb states indicating the suppression of amplitude modulation at the cavity roundtrip frequency due to the fast gain dynamics. \textbf{b} The reconstructed time domain signal does not show the generation of short pulses. This is because the intermodal difference phases in (a) correspond to the group delay. Due to the phase balance state, all parts of the spectrum arrive at a different time during on cavity roundtrip time. Nevertheless, two spikes of large output intensity are visible. This is caused by the flat intermodal difference phases around 2595 cm$^{-1}$.}
\end{figure*}

\begin{figure}[h]
	\centering
	\includegraphics[width=0.55\linewidth]{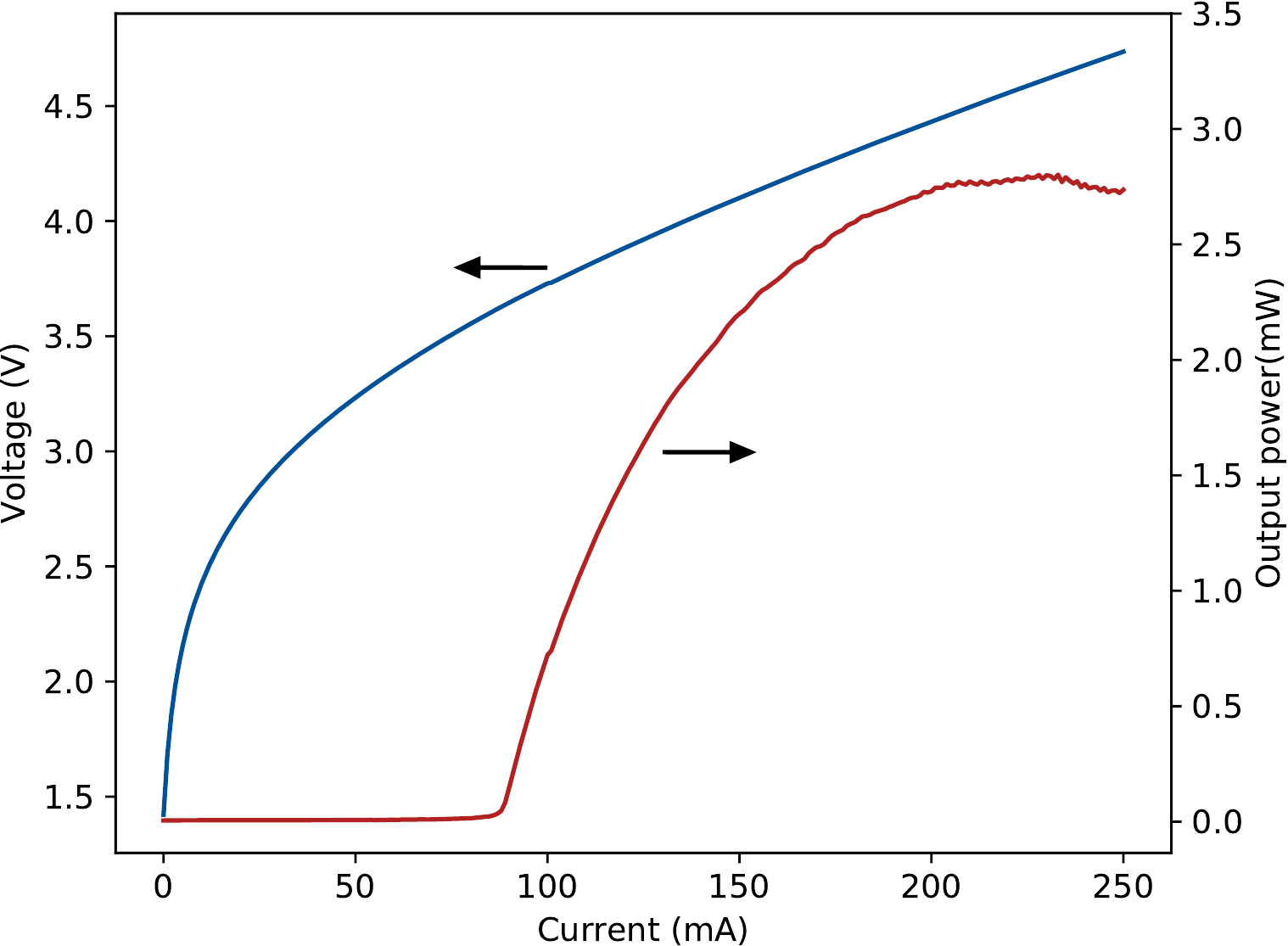}
	\caption{\textbf{Output power, current, voltage characteristics of the used ICL.} The curve was recorded with a short section bias of 3V. The maximum output power is about $2.8\,$mW and was measure with a calibrated thermal detector placed close to the laser facet. No correction factors where used.}
\end{figure}

\begin{figure}[h]
	\centering
	\includegraphics[width=0.6\linewidth]{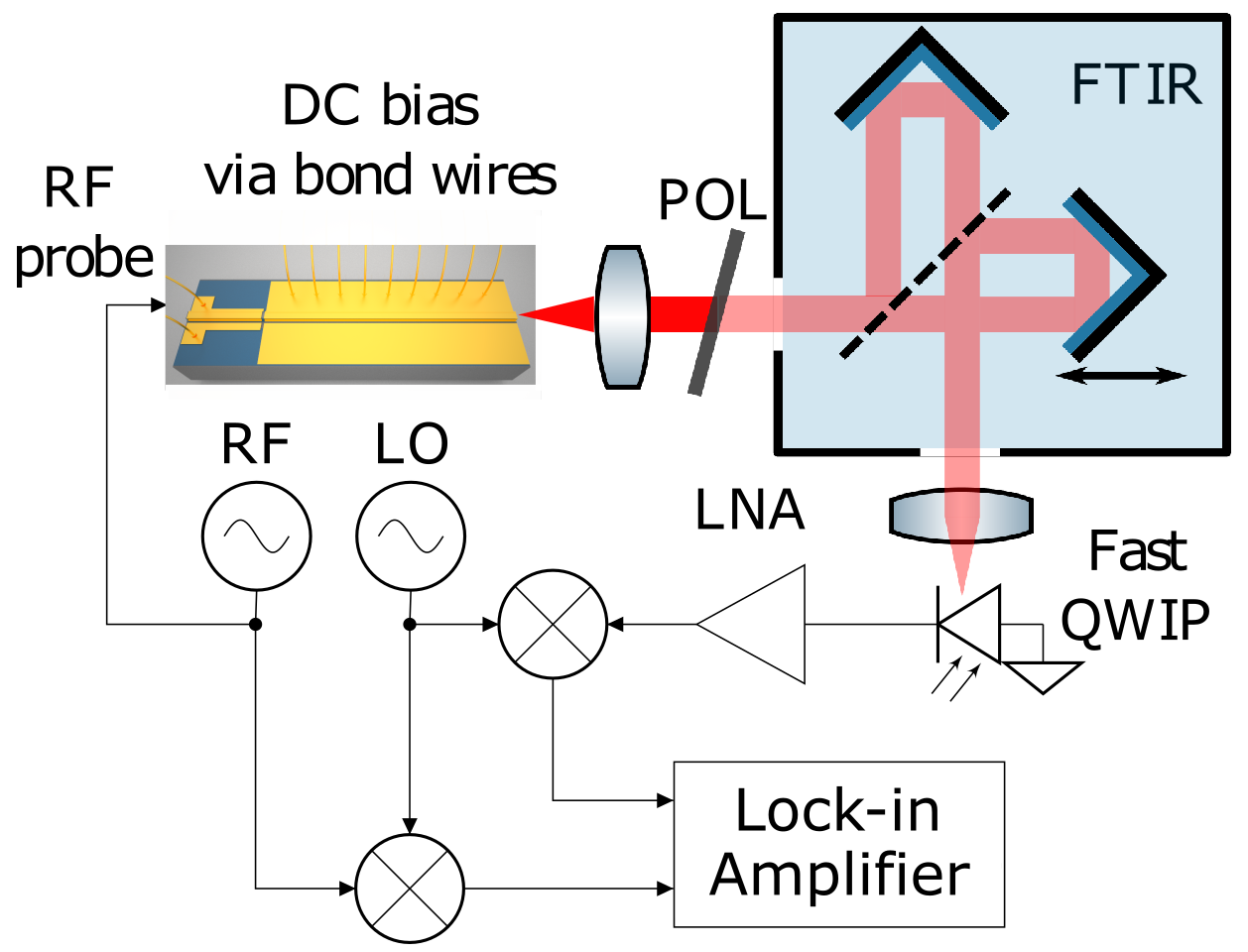}
	\caption{\textbf{Schematic drawing of the SWIFTS setup.} A RF oscillator is used to electrically lock the ICL resulting in the generation of a frequency comb. The light emitted by the ICL is attenuated by a polarizer (POL), shined through a FTIR spectrometer and detected by a fast QWIP ($\approx$ 8 GHz bandwidth). The DC intensity spectrum is obtained by recording the DC current of the QWIP. In order to obtain the SWIFTS spectra, the optical beatnote detected by the QWIP is amplified by a low-noise-amplifier (LNA) and mixed down to $\approx$ 40 MHz using a second RF oscillator. A lock-in amplifier records both SWIFTS quadrature interferograms using the Helium-Neon trigger of the FTIR spectrometer. Due to the narrowband filter of the lock-in amplifier, only the light beating at the frequency of the first RF oscillator, that is used for injection locking, contributes to the SWIFTS interferograms. In this way it is possible to judge whether the mode-spacing frequency comb is locked to the frequency of the RF oscillator.}
\end{figure}

\begin{figure}[h]
	\centering
	\includegraphics[width=0.6\linewidth]{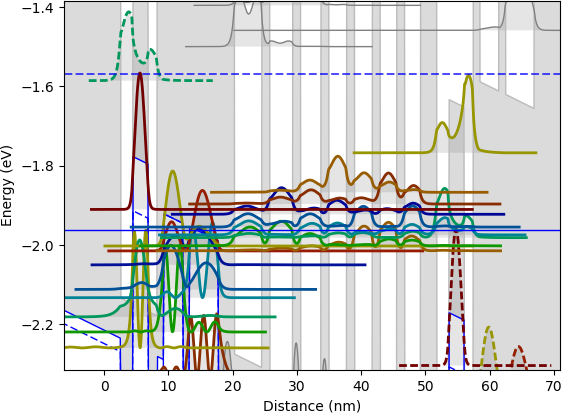}
	\caption{\textbf{Simulated bandstructure of the used ICL.} The probability densities of the envelopes were calculated using a home-build multi-band k$\cdot$p solver and are plotted at the gamma point. Most ICLs have a very similar design.}
\end{figure}